\documentclass[aps,prb,showpacs,floatfix,twocolumn,byrevtex]{revtex4-1}
\pdfoutput=1
\usepackage{amsmath}
\usepackage{amssymb}
\usepackage{amstext}
\usepackage{amsopn}
\usepackage{amsfonts}
\usepackage{amsxtra}
\usepackage{color}
\usepackage{dcolumn}
\usepackage{graphicx}
\usepackage{hyperref}
\usepackage{bm}
\begin{document}

%
% The title and the list of authors
%
\title{Effective medium approximation and the complex optical properties of the inhomogeneous
  superconductor K$_{\mathbf{0.8}}$Fe$_{\mathbf{2-y}}$Se$_{\mathbf{2}}$}
\author{C. C. Homes}
\email{homes@bnl.gov}
\author{Z. J. Xu}
\author{J. S. Wen}
\author{G. D. Gu}
\affiliation{Condensed Matter Physics and Materials Science Department,
  Brookhaven National Laboratory, Upton, New York 11973, USA}%
\date{\today}

%
% The abstract goes here
%
\begin{abstract}
The in-plane optical properties of the inhomogeneous iron-chalcogenide superconductor
K$_{0.8}$Fe$_{2-y}$Se$_2$ with a critical temperature $T_c = 31$~K have been modeled in
the normal state using the Bruggeman effective medium approximation for metallic
inclusions in an insulating matrix.  The volume fraction for the inclusions is estimated
to be $\simeq 10${\%}; however, they appear to be highly distorted, suggesting a
filamentary network of conducting regions joined through weak links.
The value for the plasma frequency $\omega_{p,D}$ in the inclusions is much larger
than the volume average, which when considered with the reasonably low values for
the scattering rate $1/\tau_D$, suggests that the transport in the grains is always
metallic.  Estimates for the dc conductivity $\sigma_{dc}$ and the superfluid density
$\rho_{s0}$ in the grains places the inclusions on the universal scaling line $\rho_{s0}/8
\simeq 4.4 \sigma_{dc}\,T_c$ close to other homogeneous iron-based superconductors.
\end{abstract}
%
%  PACS numbers
%  63.20.-e  Phonons in crystal lattices
%  89.75.Da Systems obeying scaling laws
%  72.15.-v Electronic conduction in metals and alloys
%  72.80.-r Conductivity of specific materials
%
%  74.25.Bt SC: Thermodynamic properties
%  74.25.Gz SC: Optical properties
%  74.25.Ha SC: Magnetic properties
%  74.25.Nf SC: Response to electromagnetic fields
%  74.70.-b SC:	Superconducting materials other than cuprates
%  74.72.Bk SC: Y-based cuprates
%  74.81.-g SC:	Inhomogeneous superconductors and superconducting systems, including electronic inhomogeneities
%
%  77.22.Ch  Permittivity (dielectric function)
%  78.20.-e  Optical properties of bulk materials and thin films
%  78.30.-j  Infrared and Raman spectra
%
\maketitle
%
% The main body of the text
%
% Introduction
%
\section{Introduction}
The surprising discovery of superconductivity in the iron-pnictide and
iron-chalcogenide compounds with relatively high critical temperatures ($T_c$'s)
has generated a great deal of interest in these materials.\cite{johnston10}
The minimal electronic structure is characterized by hole and electron
pockets at the center and corners of the Brillouin zone, respectively.\cite{raghu08}
It has been proposed that the scattering between the electron and hole pockets
forms the basis of a spin-fluctuation pairing mechanism.\cite{chubukov12}  For this
reason, the discovery of superconductivity in K$_{0.8}$Fe$_{2-y}$Se$_2$ was of
particular interest because the hole pocket in this material is absent,\cite{qian11,x-wang11,y-zhang11}
yet a relatively high $T_c \simeq 31$~K is observed,\cite{guo10} as opposed to the
hole-doped analog KFe$_2$As$_2$, which has a dramatically-reduced $T_c
\simeq 3$~K.\cite{chen09,sato09}  This suggests that the pairing mechanism
may not be a settled issue in these materials.\cite{f-wang11}
A further complication in understanding the physical properties of
K$_{0.8}$Fe$_{2-y}$Se$_2$ arises from the growing body of evidence that suggests
that this material is not homogeneous, but instead consists of non-magnetic
metallic (superconducting) inclusions in a magnetic, insulating
matrix.\cite{z-wang11,shen11,ricci11a,ricci11b,liu11,a-zhang11,ksenofontov11,simonelli12,
li12a,texier12,z-wang12,li12b,shermadini12,charnukha12b,c-wang12}
The optical properties of K$_{0.8}$Fe$_{2-y}$Se$_2$, and the related
Rb$_2$Fe$_4$Se$_5$ material, have been investigated in some detail and
also support the conclusion that these materials are inhomogeneous.\cite{c-wang12,
charnukha12a,yuan12,homes12}  The phase-separated nature of these materials
complicates the optical determination of the complex dielectric function,
which is by nature a volume-averaging technique.  However, a recent
study of K$_{0.8}$Fe$_{2-y}$Se$_2$ by Wang {\em et al.}\cite{c-wang12}
noted that the optical properties of this material could be described quite
well using an effective medium theory for the dielectric function which
consists of separate contributions from metallic inclusions embedded in
an insulating matrix.\cite{stroud75,walker90,brosseau02,carr85}

%
% In this work...
%
Our original study of the optical properties of K$_{0.8}$Fe$_{2-y}$Se$_{2}$
noted that the normal and superconducting state properties both indicated that
this material was inhomogeneous, and that the superconductivity was due to
Josephson coupling between the superconducting regions.\cite{homes12}  In view
of the phase-separated nature of this material, the application of an effective
medium theory to our optical data is a necessary next step in modeling
the optical properties.  In this work we apply the Bruggeman effective-medium
approximation dielectric function\cite{carr85} to the normal-state optical
properties of K$_{0.8}$Fe$_{2-y}$Se$_{2}$.  The metallic inclusions appear to
comprise about 10\% of the total sample volume, resulting in a Drude plasma
frequency that is significantly higher than the volume-averaged
value\cite{homes12} but is still much smaller that the values observed in
other (homogeneous) iron-based superconductors,\cite{yang09,wu09,hu09,lobo10,
tu10,barisic10,homes10,homes11} unless volume fractions of less than 1\% are
considered.  Interestingly, and in agreement with another recent study of
the optical properties of this material,\cite{c-wang12} the EMA can not be
applied to the data successfully without assuming that the inclusions are
extremely distorted, suggesting the formation of filamentary conducting
networks.\cite{z-wang12,hu12}  The estimated superconducting plasma frequency
of the inclusions is again much larger than the volume-averaged value, but
still significantly smaller than the values determined in other iron-based
superconductors.
The volume-averaged values for the dc conductivity (measured just above
$T_c$) and the superfluid density placed this material on the universal
scaling for the cuprate materials, albeit in a region associated with
Josephson coupling.\cite{homes12}  In contrast, the inferred superfluid
density of the metallic (superconducting) inclusions falls on the same
scaling line, but in a region associated with coherent transport and
conventional superconductivity.

%
% EMA
%
\section{Method}
There are two general theories of an effective medium.  The first
is the Maxwell Garnet dielectric function, which considers a dilute
system of inclusions.  A difficulty with this approach is that it is
asymmetric with respect to the inclusions and the matrix.
The second is the Bruggeman effective medium approximation (EMA) which is
symmetric with respect to the inclusions and the matrix and is not
restricted to any particular range of concentrations.  Another advantage
of the EMA dielectric function is that it correctly predicts the
percolation threshold for spherical grains.\cite{carr85}

%
% Figure 1 - Reflectance and unit cell.
%
\begin{figure}[tb]
\includegraphics[width=0.72\columnwidth,angle=270]{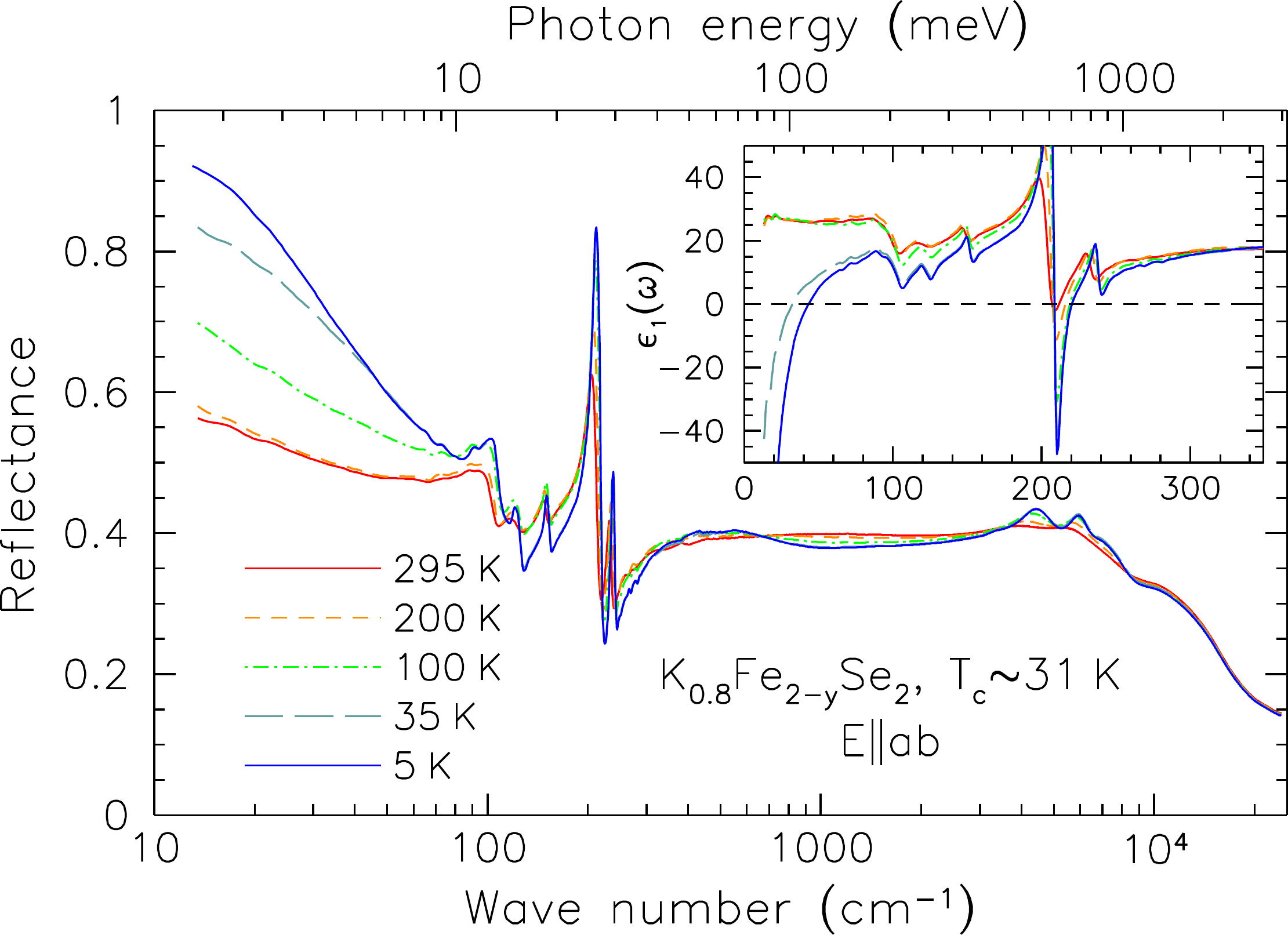}
\caption{The absolute reflectance over a wide frequency range for a cleaved
single crystal of K$_{0.8}$Fe$_{2-y}$Se$_2$ for light polarized in the
{\em a-b} planes at several temperatures above and below $T_c$.
Inset:  The temperature dependence of the real part of the dielectric
function in the far-infrared region.}
\label{fig:reflec}
\end{figure}

Because the metallic inclusions may represent a large volume of the
sample,\cite{c-wang12} we have chosen the Bruggeman EMA dielectric function.
For inclusions with complex dielectric function $\tilde\epsilon_a$ with
a volume fraction $f$ in a matrix $\tilde\epsilon_b$, the EMA dielectric
function $\tilde\epsilon$ is the root of the quadratic
expression\cite{stroud75,walker90,brosseau02,carr85}
\begin{equation}
  f{{\tilde\epsilon_a-\tilde\epsilon}\over{\tilde\epsilon_a+\phi_c\tilde\epsilon}}+
  \left(1-f\right) {{\tilde\epsilon_b-\tilde\epsilon}\over{\tilde\epsilon_b+\phi_c\tilde\epsilon}} = 0
\end{equation}
where the physical solution is the one that has ${\rm Im}(\tilde\epsilon) > 0$.  Here
$\phi_c = (1-g_c)/g_c$, where $g_c$ is the depolarization factor for a spheroid
\begin{equation}
  g_c = {{1-e_c^2}\over{e_c^2}} \left[ {1\over{e_c}} \tanh^{-1}(e_c)-1 \right].
\end{equation}
For a spheroid with figure axis length $c$ and transverse axis length $a$, the
eccentricity of a spheroid is $e_c = \sqrt{1-a^2/c^2}$.
%
% of an oblate ($c<a$) and prolate ($c>a$) spheroid is $e_o = \sqrt{a^2/c^2-1}$
% and $e_c = \sqrt{1-a^2/c^2}$ , respectively.\cite{carr85}
%
The EMA dielectric function is fit to the reflectance using a non-linear least squares
approach; the reflectance is chosen because it is a combination of the both the real
and imaginary parts of the dielectric function, as opposed to the real part of the
conductivity which depends only on the imaginary part of $\tilde\epsilon$.
The temperature dependence of the reflectance for light polarized in
the planes of a single crystal of K$_{0.8}$Fe$_{2-y}$Se$_{2}$ with
$T_c = 31$~K was determined in a previous study\cite{homes12} and is
reproduced in Fig.~\ref{fig:reflec}, with the real part of the dielectric
function shown in the inset.  Despite being a volume-averaged measurement, at
low temperature the real part of the dielectric function falls below zero at
low frequency, indicating a weakly metallic state.
The inclusions are assumed to be metallic in the normal state (superconducting
below $T_c$), with a complex dielectric function that may be described by a
simple Drude model
%
% Drude part
%
\begin{equation}
  \tilde\epsilon_a(\omega) = \epsilon_\infty -
    {{\omega_{p,D}^2}\over{\omega^2+i\omega/\tau_D}}
\end{equation}
where $\epsilon_\infty$ is the real part of the dielectric function at high
frequency, $\omega_{p,D}^2 = 4\pi ne^2/m^\ast$ and $1/\tau_D$ are the square
of the plasma frequency and scattering rate for the delocalized (Drude)
carriers, respectively, and $m^\ast$ is an effective mass.  The matrix is
assumed to be insulating with a complex dielectric function consisting only
of Lorentz oscillators
%
% Lorentz part
%
\begin{equation}
  \tilde\epsilon_b(\omega) = \epsilon_\infty +
    \sum_j {{\Omega_j^2}\over{\omega_j^2 - \omega^2 - i\omega\gamma_j}},
\end{equation}
where $\omega_j$, $\gamma_j$ and $\Omega_j$ are the position, width, and oscillator
strength of the $j$th vibration.  In addition to $\omega_{p,D}$ and $1/\tau_D$ in
$\tilde\epsilon_a$, and the stronger oscillators in $\tilde\epsilon_b$, $f$ and
$\phi_c$ are both allowed to vary; $\epsilon_\infty$ is also fit, but is assumed
to be the same in both $\tilde\epsilon_a$ and $\tilde\epsilon_b$.  (It should be
noted that in a metallic system the optical properties at low frequency are
largely independent of $\epsilon_\infty$.)
From the EMA dielectric function, at normal incidence $\tilde{r}= ({\sqrt{\tilde\epsilon}-1}) /
(\sqrt{\tilde\epsilon}+1)$
%
%\begin{equation}
% \tilde{r}= {{\sqrt{\tilde\epsilon}-1} \over {\sqrt{\tilde\epsilon}+1}}
%\end{equation}
%
and $R = \tilde{r}\tilde{r}^*$.  The complex conductivity is $\tilde\sigma(\omega)
= \sigma_1 +i\sigma_2 = -i\omega [\tilde\epsilon(\omega) -
\epsilon_\infty ]/4\pi$.

\begin{figure}[t]
\includegraphics[width=0.95\columnwidth]{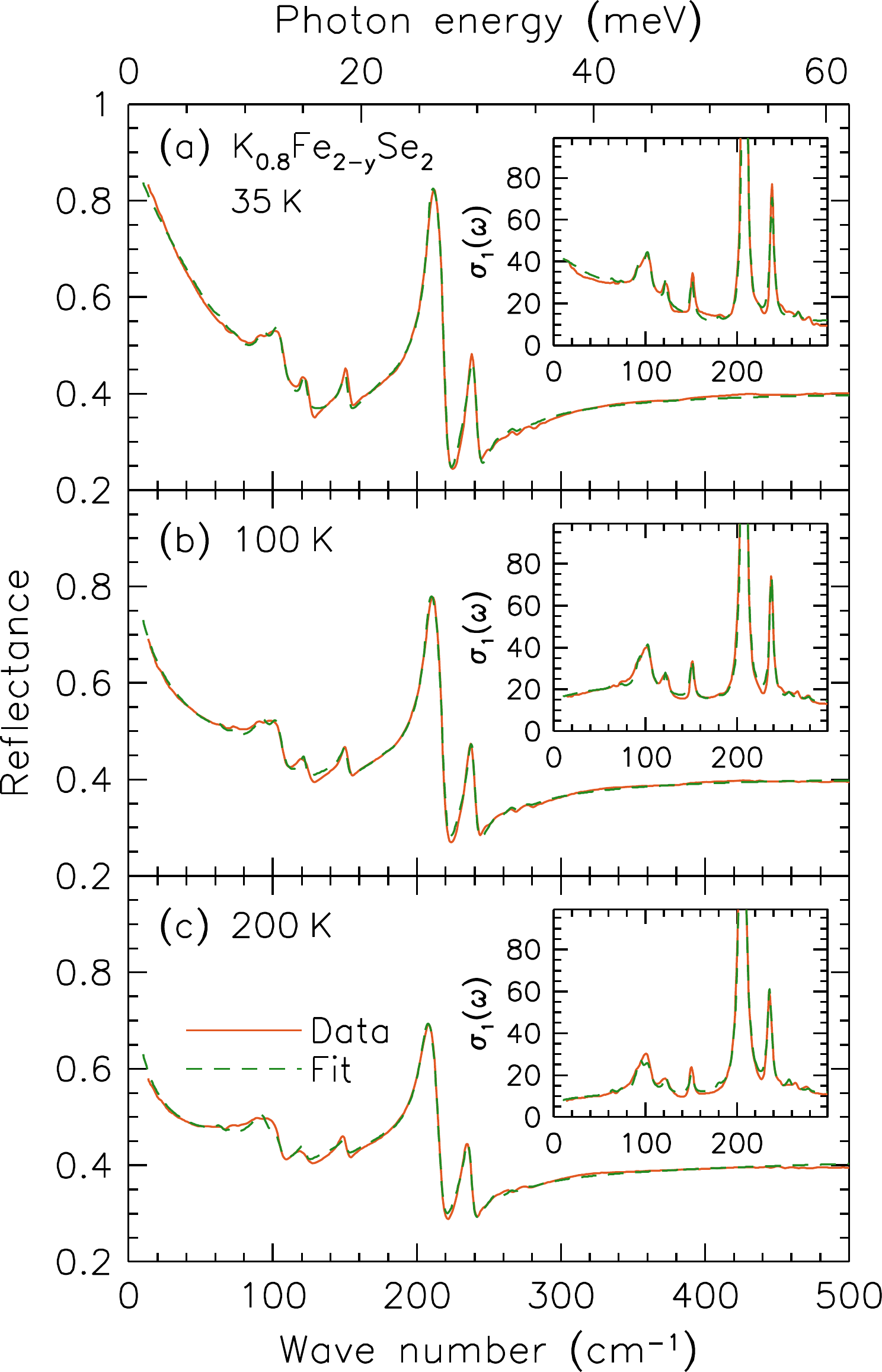}
\caption{The results of the fits of the Bruggeman EMA dielectric function
to the reflectance of K$_{0.8}$Fe$_{2-y}$Se$_2$ in the far-infrared region
for light polarized in the {\em a-b} planes at (a) 35~K, (b) 100~K and
(c) 200~K.
Inset: The fit compared to the real part of the conductivity.
}
\label{fig:fits}
\end{figure}

The results of the non-linear least squares fits of
the EMA dielectric function to the normal-state reflectance at 200, 100 and 35~K is
shown in Fig.~\ref{fig:fits}, as well as a comparison of the experimentally-determined
optical conductivity with the calculated EMA result; the agreement with experiment is
excellent.
It has been previously remarked that the quality of the EMA fits is contingent on
the size of the metallic inclusions being much smaller than the wavelength of
the radiation in the solid.\cite{c-wang12}  Given the average index of refraction
$n \simeq 4$ in this material\cite{homes12} and frequency interval of the fit, a
very rough value of the size of the inclusions is that they are no larger than
about a micron, which is in good agreement with other estimates.\cite{c-wang12,speller12}

%
% Figure 2
%

%
%n Results and discussion
%
\section{Results and Discussion}
%
% Metallic region
%
\subsection{Volume fraction}
The EMA fit to the reflectance at 35~K is shown in Fig.~\ref{fig:fits}(a)
with the comparison to the conductivity shown in the inset; the
fitted parameters are $f\simeq 0.11$ and $\phi_c \simeq 30$ with the Drude
parameters $\omega_{p,D} \simeq 1320$~cm$^{-1}$ and $1/\tau_D \simeq 57$~cm$^{-1}$.
For the fits at 100 and 200~K shown in Figs.~\ref{fig:fits}(b) and (c), $f$,
$\phi_c$ and $\omega_{p,D}$ are fixed and the scattering rate is allowed to vary,
returning $1/\tau_D \simeq 126$ and 144~cm$^{-1}$, respectively.  (Note that the
optical properties at 200 and 295~K are almost identical.\cite{homes12})  The
estimate that only about 10\% of the sample is conducting (superconducting)
is in good agreement with recent M\"{o}ssbauer,\cite{ksenofontov11} NMR,\cite{texier12}
and bulk muon-spin rotation ($\mu$SR)\cite{c-wang12,shermadini12,charnukha12b}
studies.
%
%
% Insulating region
%
The strengths of the fitted Lorentz oscillators are nearly identical to the
previously determined values,\cite{homes12} with $\epsilon_\infty \simeq 4.6$
(the same value used for the metallic inclusions).  Given that the insulating
matrix accounts for about 90\% of the sample volume, it is not surprising that
the vibrational parameters should remain essentially unchanged.
Previous Drude-Lorentz fits to the volume-averaged optical conductivity at
35~K yielded $\omega_{p,D}^{vol} \simeq 430$~cm$^{-1}$ and $1/\tau_D^{vol}
\simeq 70$~cm$^{-1}$;\cite{homes12} this result is consistent with the
typically low values of $\omega_{p,D}^{vol}$ observed in these
materials.\cite{yuan12,charnukha12a,c-wang12}  If we attribute this
average plasma frequency to the fraction $f$ of the sample that is
metallic, then
\begin{equation}
  {{\omega_{p,D}}\over{\omega_{p,D}^{vol}}} = {1\over{\sqrt{f}}}.
\end{equation}
For $f\simeq 0.11$, $\omega_{p,D} = \omega_{p,D}^{vol}/\sqrt{f} \simeq
1300$~cm$^{-1}$; this value is almost identical to the fitted EMA value
of $\omega_{p,D} \simeq 1320$~cm$^{-1}$.
%
% Still have a small value for w_pD
%
The value for $\omega_{p,D}$ is still rather small when compared with
values of $\omega_{p,D} \simeq 7000 - 14\,000$~cm$^{-1}$ observed in other
iron-based superconductors.\cite{yang09,wu09,hu09,lobo10,barisic10,tu10,homes10}
Indeed, for $\omega_{p,D}$ to rival these values would require a
volume fraction of less than 1\%.  Setting the fraction of metallic
inclusions to $f=0.005$ yields the EMA
fitted value of $\omega_{p,D} \simeq 6620$~cm$^{-1}$ (close to the value
of 6080~cm$^{-1}$ based on the volume average), and $1/\tau_D \simeq
38$~cm$^{-1}$; however, the fitted value $\phi_c \gtrsim 200$
is quite large and the over quality of the fit has decreased significantly.
In either case, the temperature dependence of the volume-averaged conductivity
was originally described as incoherent at room temperature with a large scattering
rate that decreases rapidly with temperature resulting in a crossover to coherent
behavior at low temperature.  However, assuming $f\simeq 0.1$, the EMA values for
$1/\tau_D$ suggest that the transport in the metallic regions is always coherent.

%
% Distorted shapes
%
The value for $\phi_c \simeq 30$ yields the rather small value for the
depolarization factor $g_c \simeq 0.032$ which corresponds to an eccentricity
$e_o \simeq 0.93$ for an oblate spheroid or $e_c \simeq 0.99$ in a prolate
spheroid; both cases correspond to highly distorted shapes.  This condition
becomes even more severe for larger values of $\phi_c$.  The layered nature
of these materials\cite{bao11,bacsa11,zavalij11} and the anisotropic transport
properties\cite{h-wang11} suggests that these distorted shapes overlap or are
joined through weak links to form a conducting pathways through the solid,
resulting in a predominantly two-dimensional filamentary network, or a
superconducting aerogel.\cite{z-wang12,hu12}

%
% Isolated inclusions?
%
In this approach we have assumed that the highly-distorted inclusions overlap
to some degree.  However, it might also be possible that a large number of
inclusions may be isolated , in which case for spherical particles with
$\epsilon_\infty=1$ for both the inclusions and the matrix, the effective
dielectric function would experience a resonance at $\omega_0 = \omega_{p,D}
\sqrt{(1-f)/3} \simeq 720$~cm$^{-1}$ (Maxwell Garnet theory\cite{carr85}).
However, the values $\epsilon_\infty = 4.6$ and $\phi_c\simeq 30$ dramatically
lower this resonance, $\omega_0 \lesssim 100$~cm$^{-1}$.  The absence of such a
feature in our results suggests that either a continuous distribution of shapes has
rendered this feature too broad and weak to be observed, or that there are simply
very few isolated inclusions.

%
% Energy scales
%
\subsection{Energy scales}
In the iron-chalcogenide superconductors, the energy scales for the isotropic
superconducting energy gaps that are observed in angle-resolved photoemission
spectroscopy (ARPES) to open on the hole and electron pockets below $T_c$ are
usually in excellent agreement with the optical gaps observed in the conductivity
that develop in the superconducting state.\cite{homes10,miao12}  However, these
two energy scales appear to be very different in K$_{0.8}$Fe$_{2-y}$Se$_2$.
While the ARPES estimate of the isotropic optical gap is $2\Delta \simeq 16 - 20$~meV
($\simeq 130 - 160$~cm$^{-1}$),\cite{qian11,x-wang11,y-zhang11} the reflectance
(and the conductivity) indicates that the energy scale associated with the
superconductivity in this material is much smaller, $\simeq 8$~meV.
This difference originates from the inhomogeneous nature of this material.
ARPES is insensitive to the insulating matrix and directly probes the formation of a
superconducting gap in the metallic (superconducting) inclusions, while the optical properties are
a volume-averaging technique which will be sensitive to the Josephson coupling between
the superconducting regions.  In such a Josephson coupled system, changes in the reflectance
(for instance) will occur not at $2\Delta$ but at the renormalized superconducting
plasma frequency, $\tilde\omega_{p,S} = \omega_{p,S}/\sqrt{\epsilon_{\rm FIR}}$, or
at the average value for a distribution of frequencies.\cite{yuan12}
Given $\omega_{p,S}\simeq 220$~cm$^{-1}$ and $\epsilon_{\rm FIR}\simeq 18$ at
50~meV (inset of Fig.~\ref{fig:reflec}), then $\tilde\omega_{p,S}\simeq 52$~cm$^{-1}$
or $\simeq 6.5$~meV, which is very close to the changes in the optical properties
observed to occur below $\simeq 8$~meV.
Thus, due to the inhomogeneous nature of this superconductor, optics and ARPES
probe two different quantities, $\tilde\omega_{p,S}$ and $\Delta$, respectively.

%
% SC response: calculate wps from relation above
%
\subsection{Parameter scaling}
It has been pointed out that a number of the iron-based
superconductors\cite{wu10} fall on the scaling relation initially
observed for the cuprate superconductors,\cite{homes04,homes05a,homes05b}
$\rho_{s0}/8 \simeq 4.4\,\sigma_{dc}\,T_c$, where the superfluid density is
$\rho_{s0} \equiv \omega_{p,S}^2$.
In our previous optical study of K$_{0.8}$Fe$_{2-y}$Se$_2$ the volume averaged
value for the superconducting plasma frequency was determined to be
$\omega_{p,S}^{vol}\simeq 220$~cm$^{-1}$ (Ref.~\onlinecite{homes12}).
While this value is quite small, this material does indeed fall on the
universal scaling line; however, it does so in a region associated with the
response along the {\em c} axis in the cuprates where the superconductivity
is due to Josephson coupling between the copper-oxygen planes.  From this it
was concluded that the superconductivity was due to the Josephson coupling of
discrete superconducting regions and that the material constituted a Josephson
phase.\cite{imry08,imry12}

\begin{figure}[t]
%\centerline{\includegraphics[width=2.8in]{figure3.eps}}
\includegraphics[width=0.95\columnwidth]{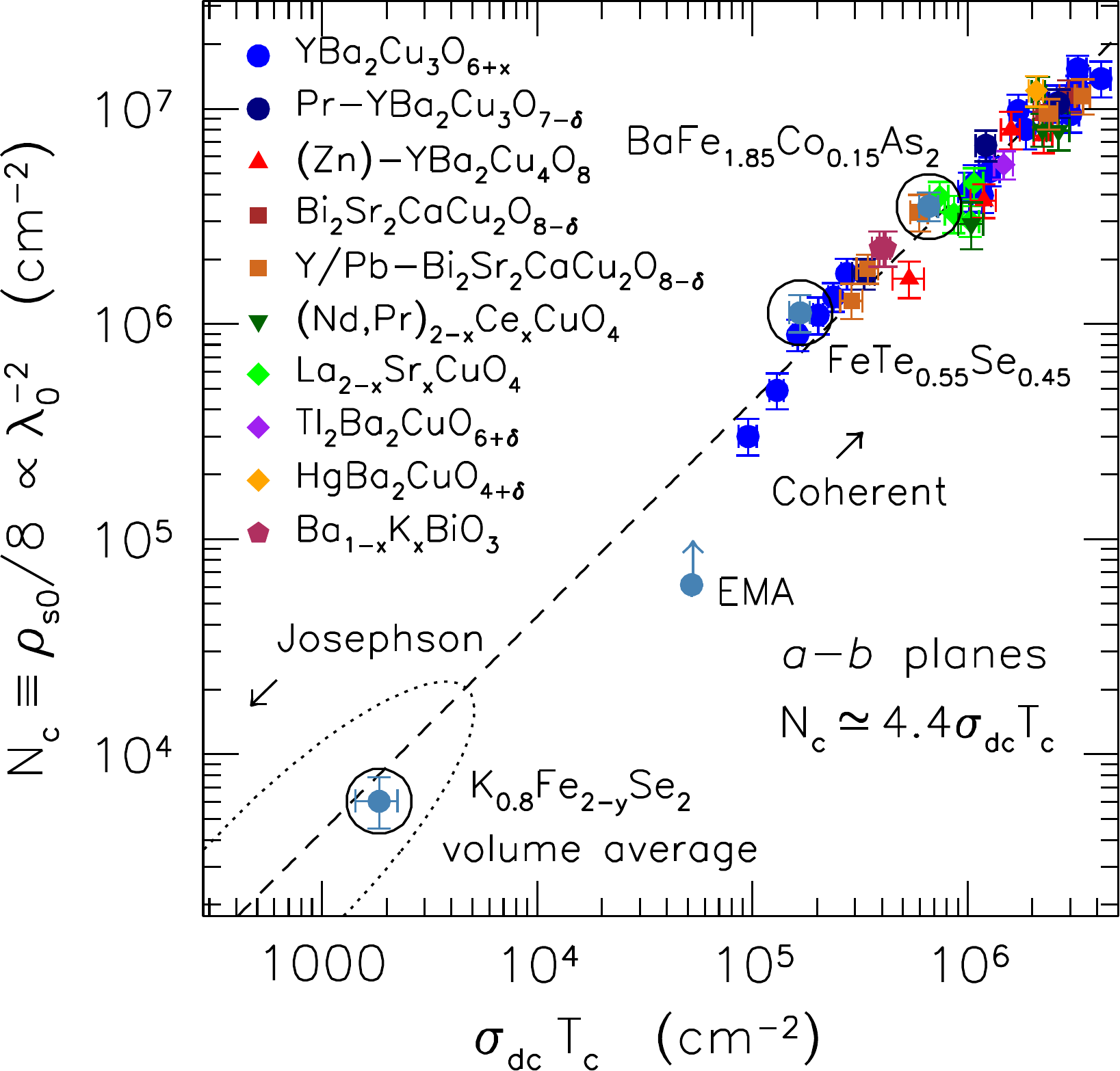}
\caption{The log-log plot of the spectral weight of the superfluid
 density $N_c \equiv \rho_{s0}/8$ vs $\sigma_{dc}\,T_c$ in the
 {\em a-b} planes for a variety of cuprate superconductors, as well
 as several iron-based supercondcutors, compared with the volume
 average and EMA results for K$_{0.8}$Fe$_{2-y}$Se$_2$.
 The dashed line corresponds to the general result for the cuprates
 $\rho_{s0}/8 \simeq 4.4 \sigma_{dc} T_c$, while the dotted line
 denotes the region of the scaling relation typically associated
 with Josephson coupling along the {\em c} axis.  While the volume
 average result signalled a Josephson phase, the EMA result
 now lies very close to the coherent regime.}
\label{fig:scale}
\end{figure}

%
% SC plasma frequency
%
The superconducting plasma frequency for the inclusions with $f=0.1$ may be
estimated below $T_c$ using $\omega_{p,S} = \omega_{p,S}^{vol}/\sqrt{f}
\simeq 700$~cm$^{-1}$, which corresponds to an effective penetration depth of
$\lambda_{eff} \simeq 2.2$~$\mu$m.
We note that the EMA model yields a lower value for the normal-state scattering
rate $1/\tau_D \simeq 57$~cm$^{-1}$ at 35~K than the volume average, $1/\tau_D^{vol}
\simeq 70$~cm$^{-1}$.  This smaller value for $1/\tau_D$ results in the condition
$1/\tau_D < 2\Delta$.  This would normally imply that more spectral weight (the
area under the conductivity curve) associated with the free carriers lies in the
gap and thus more spectral weight should be transferred to the condensate.\cite{homes05a}
In the volume average case, only about 25\% of the free carriers collapse
into the condensate; the EMA result implies that the estimate of $\omega_{p,S}
\simeq 700$~cm$^{-1}$ likely represents a minimum value.
%
% Scattering rate
%
However, the issue of the scattering rate itself is somewhat complicated.
Despite the fact that this material has only electron pockets, it has been
proposed that the scattering rate is anisotropic;\cite{kemper11} this has
resulted in some workers adopting a two-component model with large and
small scattering rates.\cite{charnukha12a,c-wang12}  In the EMA fits used
here, only a single component has been employed. Therefore, if there is
in fact a distribution of scattering rates, the fitted EMA value will
represent an average value.  As a result, there is some uncertainly
attached to the value of $1/\tau_D$.  With this caveat in place, the
EMA fit to the reflectance just above $T_c$ at 35~K may be used to estimate
the dc conductivity of the metallic inclusions, $\sigma_{dc} =
\omega_{p,D}^2\,\tau_D/60 \simeq 510$~$\Omega^{-1}$cm$^{-1}$.
The values for $\omega_{p,S}$ and $\sigma_{dc}$ once again place
this material close to the scaling line, but now the material falls
very close to the other iron-chalcogenide superconductors, as shown
in Fig.~\ref{fig:scale}.
%
% Figure 3
%

While $\omega_{p,S}$ is significantly larger than the volume-averaged value,
it is still almost an order of magnitude smaller than $\mu$SR\cite{c-wang12,
shermadini12} and NMR\cite{torchetti11} estimates, although adopting a
smaller value for the volume fraction $f$ negates this difference.  On the other
hand, the value of $\lambda_{eff} \simeq 2.2$~$\mu$m in the superconducting
regions of K$_{0.8}$Fe$_{2-y}$Se$_2$ is in surprisingly good agreement with
the in-plane optical estimate of $\lambda\simeq 2$~$\mu$m in Rb$_2$Fe$_4$Se$_5$
using an EMA approach.\cite{charnukha12b}

%
% Conclusions
%
\section{Conclusions}
The complex optical properties of K$_{0.8}$Fe$_{2-y}$Se$_2$ in the normal
state have been modeled using the Bruggeman EMA.  The volume fraction of
the metallic inclusion is estimated to be $f \simeq 0.1$; however, the EMA
can only be successfully fit to the data if the inclusions are highly
distorted, suggesting a filamentary network of conducting regions joined
through weak links.
The plasma frequency in the metallic inclusions is therefore considerably
larger than the volume-averaged value, $\omega_{p,D} > \omega_{p,D}^{vol}$;
however, $\omega_{p,D} \simeq 1320$~cm$^{-1}$ is still much smaller
than the values for the plasma frequency observed in other (homogeneous)
iron-based superconductors, as is the estimate of $\omega_{p,S} \simeq
700$~cm$^{-1}$ (unless volume fractions of less than 1\% are considered).
The reasonably small values for $1/\tau_D \simeq 60 - 140$~cm$^{-1}$ returned
by the EMA fits suggests that the transport in the metallic regions is
always coherent, and that there is no crossover from incoherent behavior as
the temperature is lowered.
The inferred $\sigma_{dc} \simeq 510$~$\Omega^{-1}$cm$^{-1}$ just above
$T_c$ and the estimated lower bound of $\rho_{s0} \simeq 4.9\times 10^5$~cm$^{-2}$
for the metallic (superconducting) inclusions shifts this material away from the region
on the scaling line associated with Josephson coupling to a region where the
majority of (homogeneous) iron-based superconductors are observed to lie.

%
% Acknowledgements...
%
\begin{acknowledgements}
We would like to thank A. Akrap, G. L. Carr, A. Charnukha, and
D. van der Marel for useful discussions.
Research supported by the U.S. Department of Energy, Office of
Basic Energy Sciences, Division of Materials Sciences and Engineering
under Contract No. DE-AC02-98CH10886. Z.~X. and J.~W. are supported
by the Center for Emergent Superconductivity, an Energy
Frontier Research Consortium supported by the Office of
Basic Energy Science of the Department of Energy
\end{acknowledgements}
%

%
%%%%%%%%%%%%%%%%%%%%%%%%%%%%%%%%%%%%%%%%%%%%%%%%%%%%%%%%%%%%%%%%%%%%%%%%%%%%%%
%
% References
%
%\bibliography{kfs_ema}
%


\end{document}